\documentstyle[preprint,aps,epsfig]{revtex}
\begin{document}
\draft
\title{Dynamics of a polymer test chain in a glass forming matrix: the
Hartree approximation}
\author{M. Rehkopf$^{(a)}$,
        V.G. Rostiashvili$^{(a,b)}$ and
        T.A. Vilgis$^{(a)}$}
\address{$^{(a)}$Max-Planck-Institut f\"ur Polymerforschung, Postfach 3148,
D-55021
  Mainz, Germany}
\address{$^{(b)}$ Institute of Chemical Physics, Russian Academy of Science,
142432, Chernogolovka, Moscow region, Russia}
\date{\today}
\maketitle
classification: Physics abstracts, 05.20-36.20
\begin{abstract}
We consider the Langevin dynamics of a Gaussian test polymer chain coupled
with a surrounding matrix which can undergo the glass transition. The
Martin-Siggia-Rose generating functional method and the nonpertubative
Hartree approximation are used
to derive the generalized Rouse equation for the test chain. It is shown that
the interaction of the test chain with the surrounding matrix renormalizes the
bare friction and the spring constants of the test chain in such a way that
the memory function as well as the bending dependent elastic modulus
appear. We find that below the glass transition temperature $T_{G}$ of the
matrix the Rouse modes of the test chain can be frozen and moreover the
freezing temperatures (or the ergodicity-nonergodicity transition temperature)
$T_{c}(p)$ depends from the Rouse mode index $p$.
\end{abstract}
\section{Introduction}
It is wellknown that for relatively short polymer chains the standard Rouse
model can describe the dynamics of a melt reasonably well \cite{Doi,Ferry}.
On
the contrary, for chain length $N$ exceeding a critical length, the
entanglement length $N_{e}$, the behavior is usually described by
the reptation model \cite{Doi}.
Here we restrict ourselves to chain lengths $N<N_{e}$, i.e.
the entangled polymer dynamics will be beyond of our consideration.

The reason why in a dense melt the Rouse model provides so well dynamical
description for short chains is connected with a screening of the long-range
hydrodynamic as well as the excluded volume interactions.
As a result the fluctuations of the chain variables are Gaussian.
But there are further essential questions: How does the bare monomeric
friction coefficient $\xi_{0}$ and the entropic elastic modulus
$\varepsilon$ (which are simple
input parameters of the standard Rouse model) change due to the
interactions of the test chain and the surrounding matix?
Why does such a simple model work
so well for describing short chain melts? Obviously, the corresponding
answers cannot be given by the Rouse model, which describes only
the dynamics of connected Gaussian springs without further
interactions.

On the other hand, at relatively low temperatures close to the glass
transition of the surrounding matrix the deviations from the standard Rouse
behavior will be definitely more pronounced. For example, Monte Carlo (MC)
studies of the bond fluctuation model at low temperatures
(but still above the
temperature region where possibly the glass transition mode coupling theory
\cite{Gotze} applies) show that the Rouse modes remain
well-defined eigenmodes
of the polymer chains and the chains retain their Gaussian properties
\cite{Binder}. Nevertheless, the relaxation of the Rouse modes displays a
stretched exponential behavior rather than a pure exponential. It could
even be
expected that at temperatures below the glass transition temperature of the
matrix $T_{G}$ the Rouse modes are frozen out. In these temperature regimes
the interactions between monomers take a significant role and determine
the physical picture of the dynamics as will be shown below.

 The generalized Rouse
equation (GRE), which can be used for the investigation of the problems
mentioned above, has been derived by using projection formalism
methods and
mode coupling approximations (MCA) \cite{Hess,Schweizer1,Schweizer2}.
As a result of
projection operator formalism the time evolution of the
test chain is
expressed in terms of a frequency matrix, which is local in
time, and a memory
function contribution due to the inter-chain forces exerted on
the test chain
segments. With the assumption that the
frequency matrix term has the same form
as in the standard Rouse model (linear elasticity with the entropic modulus
$\varepsilon=3k_{b}T/l^2$) all influence of the matrix chains reduce to the
memory function contribution \cite{Hess,Schweizer1,Schweizer2}.

The
projection operator methods appears to be exact but rather formal, and to
derive
explicit results further approximations have to be made,
which can be hardly
controlled often. Therefore it is instructive
to use another alternative theoretical method to
derive the GRE. Recently, a non-pertubative variational method which is
equivalent to a selfconsistent Hartree approximation was used for the
investigation of the dynamics of manifolds \cite{Horner} and sine-Gordon model
\cite{Cule} in a random media. As a starting point the authors employed the
standard Martin-Siggia-Rose (MSR) functional integral technique
\cite{Dominicis,Bausch}. Here we follow this approach to derive a GRE and
study the dynamics of a test polymer chain in a glass forming matrix.\\
The paper is organized as follows. In section 2, we give a general
MSR-functional integral formulation for a test chain in a polymer (or
non-polymer) matrix. Under the assumption that the fluctuations of the test
chain are Gaussian the Hartree-type approximation is applied and a GRE is
finally derived. The case when the fluctuation dissipation theorem (FDT) and
the time homogenity are violated is also shortly considered. In section 3 on
the basis of the GRE some static and dynamical properties of the test chain
are discussed. In particular the theory of the test chain ergodicity breaking
(freezing) in a glassy matrix is formulated. Section 4 gives some summary and
general discussion. The appendices are devoted to some technical details of the
Hartree-type approximation.
\section{Generalized Rouse Equation (GRE)}
\subsection{MSR-functional integral approach}
Let us consider a polymer test chain with configurations characterized by
the vector function ${\bf R}(s,t)$ with s numerating the segments of the
chain, $0\le s\le N$, and time $t$. The test polymer chain moves in the melt
of the other polymers (matrix) which positions in space are specified by the
vector functions ${\bf r}^{(p)}(s,t)$, where the index $p=1,2,...,M$
numerates the different chains of the matrix. The test chain is expected to
have Gaussian statistics due to the screening of the self-interactions in a
melt \cite{Doi}. We consider the simultaneous dynamical evolution of the
${\bf R}(s,t)$ and ${\bf r}^{(p)}(s,t)$ variables assuming that the
interaction between matrix and test chain is weak.\\ The Langevin equations
for the full set of variables $\{{\bf R}(s,t),{\bf r}^{(1)}(s,t),\ldots,{\bf
r}^{(M)}(s,t)\}$ has the form
\begin{eqnarray}
\xi_{0}\frac{\partial}{\partial
t}R_{j}(s,t)&-&\varepsilon\frac{\partial^2}{\partial
s^2}R_{j}(s,t)+\frac{\delta}{\delta R_{j}(s,t)}\sum_{p=1}^{M}\int_{0}^{N}
ds' V\left({\bf R}(s,t)-{\bf
r}^{(p)}(s',t)\right)\nonumber\\&=&f_{j}(s,t)\label{R}
\end{eqnarray}
\begin{eqnarray}
\xi_{0}\frac{\partial}{\partial
t}r_{j}^{(p)}(s,t)&-&\varepsilon\frac{\partial^2}{\partial
s^2}r_{j}^{(p)}(s,t)+\frac{\delta}{\delta r_j^{(p)}(s,t)}\sum_{m=1}^M\int_0^N
ds' {\tilde V}\left({\bf r}^{(p)}(s,t)-{\bf
r}^{(m)}(s',t)\right)\nonumber\\&+&\frac{\delta}{\delta
r_j^{(p)}(s,t)}\sum_{m=1}^M\int_0^N ds' V\left({\bf r}^{(p)}-{\bf
R}(s',t)\right)={\tilde f}_j(s,t)\label{r}
\end{eqnarray}
where $\xi_0$ denotes the bare friction coefficient, $\varepsilon=3T/l^2$
the bare elastic modulus with the length of a Kuhn segment denoted by $l$,
$V(\cdots)$ and ${\tilde V}(\cdots)$ are the interaction energies of test
chain-matrix and matrix-matrix respectively, and $f_j(s,t)$, ${\tilde
f}_j(s,t)$ are the random forces with the correlator
\begin{equation}
\left<f_i(s,t)f_j(s',t'))\right>=\left<{\tilde f}_i(s,t){\tilde
f}_j(s',t')\right>=2T\xi_0\delta_{ij}\delta(s-s')\delta(t-t')\label{f}.
\end{equation}
After using the standard MSR-functional integral representation
\cite{Dominicis} for the system (\ref{R}-\ref{f}), the generating functional
(GF) takes the form
\begin{eqnarray}
&Z&\left\{\cdots\right\}=\int DR_j(s,t)D{\hat R}_j(s,t)\int\prod_{p=1}^M
Dr_j^{(p)}(s,t)D{\hat
r}_j^{(p)}(s,t)\label{GF}\\&\times&\exp\Bigg\{-A_0\left[{\bf R}(s,t),{\bf
{\hat R}}(s,t)\right]-A_1\left[{\bf r}^{(p)}(s,t),{\bf {\hat
r}}^{(p)}(s,t)\right]\nonumber\\&+&\sum_{p=1}^M\int_0^N ds\int_0^N ds'\int
dt\:i{\hat R}_j(s,t)\frac{\delta}{\delta R_j(s,t)}V\left[{\bf R}(s,t)-{\bf
r}^{(p)}(s',t)\right]\nonumber\\&+&\sum_{p=1}^M\int_0^N ds\int_0^N ds'\int
dt\:i{\hat r}^{(p)}_j(s',t)\frac{\delta}{\delta r_j^{(p)}(s',t)}V\left[{\bf
r}^{(p)}(s',t)-{\bf R}(s,t)\right]\Bigg\}\nonumber
\end{eqnarray}
where the dots represents some source fields which will be specified later and
Einstein's summation convention for repeated indices is used.
In GF (\ref{GF}) the MSR-action of the free test chain is given by
\begin{eqnarray}
A_0\left[{\bf R}(s,t),{\bf {\hat R}}(s,t)\right]&=&-\int_0^N ds\int
dt\Bigg\{i{\hat R}_j(s,t)\left[\xi_0\frac{\partial}{\partial
t}R_j(s,t)-\varepsilon\frac{\partial^2}{\partial
s^2}R_j(s,t)\right]\nonumber\\&+&T\xi_0\left[i{\bf {\hat
R}}(s,t)\right]^2\Bigg\}\qquad.
\end{eqnarray}
As we will realize later the explicit form of the full action of the medium
$A_1\left[{\bf r}^{(p)}(s,t),{\bf {\hat r}}^{(p)}(s,t)\right]$ plays no role.
In principle it could have any form and in particular, for a polymer matrix,
the following one
\begin{eqnarray}
&\:&{A_1}\left[{\bf r}^{(p)}(s,t),{\bf {\hat
r}}^{(p)}(s,t)\right]=-\sum_{p=1}^M\int_0^N ds\int dt\:i{\hat
r}_j^{(p)}(s,t)\left[\xi_0\frac{\partial}{\partial
t}r_j^{(p)}(s,t)-\varepsilon\frac{\partial^2}{\partial
s^2}r_j^{(p)}(s,t)\right]\nonumber\\&-&\sum_{p=1}^M\int_0^N ds\int dt\:i{\hat
r}_j^{(p)}(s,t)\frac{\delta}{\delta r_j^{(p)}(s,t)}\sum_{m=1}^M\int
ds'{\tilde V}\left[{\bf {\hat r}}^{(p)}(s,t)-{\bf {\hat
r}}^{(m)}(s',t)\right]\nonumber\\&+&\sum_{p=1}^M\int_0^N ds\int
dt\:T\xi_0\left[i{\hat r}_j(s,t)\right]^2
\end{eqnarray}
In order to obtain an equation of motion for the test chain one should
integrate over the matrix variables ${\bf r}^{(p)}(s,t)$ first. For this
end it is reasonable to represent GF (\ref{GF}) as
\begin{eqnarray}
Z\left\{\cdots\right\}&=&\int DR_j(s,t)D{\hat R}_j(s,t)\nonumber\\
&\times&\exp\left\{-\Xi\left[R_{j}(s,t),{\hat
      R}_{j}(s,t)\right]-A_{0}\left[{\bf R}(s,t),{\bf {\hat
        R}}(s,t)\right]\right\}\label{GF1}
\end{eqnarray}
where the influence functional $\Xi$ is given by
\begin{eqnarray}
&\:&\Xi\left[{\bf R},{\bf {\hat R}}\right]=-\ln\int\prod_{p=1}^{M}D{\bf
  r}^{(p)}(s,t)D{\bf {\hat
r}}^{(p)}(s,t)\times\label{inf}\\&\times&\exp\Bigg\{-A_{1}\left[{\bf
      r}^{(p)},{\bf {\hat r}}^{(p)}\right]\nonumber\\
&+&\sum_{p=1}^M\int_0^N ds\int_0^N ds'\int dt\:i{\hat
    R}_{j}(s,t)\frac{\delta}{\delta R_{j}(s,t)}V\left[{\bf R}(s,t)-{\bf
r}^{(p)}(s',t)\right]\nonumber\\&+&\sum_{p=1}^M\int_0^N ds\int_0^N ds'\int
dt\:i{\hat
    r}_{j}^{(p)}(s',t)\frac{\delta}{\delta r_{j}(s',t)}V\left[{\bf
      r}^{(p)}(s',t)-{\bf R}(s,t)\right]\Bigg\}\nonumber.
\end{eqnarray}
In the spirit of the mode coupling approximation (MCA) \cite{Gotze,Schweizer1} the force between the test chain and the matrix should be expressed as a bilinear product of the two subsystems densities. In order to assure this we expand the influence functional (\ref{inf} with respect to the forces $F_j=-\nabla_{j}V$ between the test chain and the matrix up to the second order. This leads to
\begin{eqnarray}
&{\Xi\left[{\bf R},{\bf {\hat R}}\right]}&=-\ln\int\prod_{p=1}^{M}D{\bf
  r}^{(p)}(s,t)D{\bf {\hat r}}^{(p)}(s,t)\Bigg\{\exp\left\{-A_{1}\left[{\bf
      r}^{p},{\bf {\hat r}}^{p}\right]\right\}\nonumber\\&+&\frac{1}{2!}\int
d^{3}rd^{3}r'\int ds\int ds'\int dt\:i{\hat
    R}_{j}(s,t)\frac{\delta}{\delta R_{j}(s,t)}V\left[{\bf R}(s,t)-{\bf
r}\right]\nonumber\\&\:&\qquad\times\int dt'\:i{\hat
    R}_{l}(s',t')\frac{\delta}{\delta R_{l}(s',t')}V\left[{\bf
      R}(s',t')-{\bf r}'\right]\left<\rho({\bf r},t)\rho({\bf
      r}',t')\right>_{1}\nonumber\\&+&\frac{1}{2!}\int
  d^{3}rd^{3}r'\int ds\int ds'\int dt\int dt'V\left[{\bf r}-{\bf
      R}(s,t)\right]V\left[{\bf r}'-{\bf
R}(s',t')\right]\nonumber\\&\;&\qquad\times\nabla_{l}\nabla_{j}'\left<\Pi_{l}({\bf
r},t)\Pi_{j}({\bf r}',t')\right>_{1}\nonumber\\&-&\frac{1}{2!}\int
  d^{3}rd^{3}r'\int ds\int ds'\int dt\int_{-\infty}^{t} dt'\:i{\hat
    R}_{j}(s,t)\frac{\delta}{\delta R_{j}(s,t)}V\left[{\bf R}(s,t)-{\bf
      r}\right]\nonumber\\
&\:&\qquad\times V\left[{\bf r}'-{\bf
      R}(s',t')\right]\nabla_{l}'\left<\rho({\bf r},t)\Pi_{l}({\bf
      r}',t')\right>_{1}\nonumber\\&+&(t\Leftrightarrow t')+{\cal
O}(F^{3})\Bigg\}\label{einf}
\end{eqnarray}
where the matrix density
\begin{equation}
\rho({\bf r},t)=\sum_{p=1}^{M}\int_{0}^{N}ds\delta\left({\bf r}-{\bf
    r}^{(p)}(s,t)\right)
\end{equation}
and the response field density
\begin{equation}
\Pi_{j}({\bf r},t)=\sum_{p=1}^{M}\int_{0}^{N}ds\:i{\bf {\hat
r}}_{j}^{(p)}(s,t)\delta\left({\bf r}-{\bf
    r}^{(p)}(s,t)\right)
\end{equation}
were introduced and $\left<\cdots\right>_{1}$ denotes cumulant averaging over
the full MSR-action $A_{1}\left[{\bf r},{\bf {\hat r}}\right]$ of the
matrix. In eq.~(\ref{einf}) the term $(t'\Leftrightarrow t)$ is the same
like the previous one but with permutated time arguments. The terms which are
linear with respect to $F_{j}$ vanishes because of the homogenity of the
system. In the Appendix A we show that because of causality the correlator
$\left<\Pi_{l}({\bf r},t)\Pi_{j}({\bf r}',t')\right>_{1}$ equals
zero \cite{Dominicis,Bausch,Ros}. Taking this into account and performing
the spatial Fourier
transformation the expression for GF (\ref{GF1}) takes the form
\begin{eqnarray}
Z\left\{\cdots\right\}&=&\int DR_j(s,t)D{\hat
  R}_j(s,t)\exp\Bigg\{-A_{0}\left[{\bf R}(s,t),{\bf {\hat
      R}}(s,t)\right]\nonumber\\&+&\frac{1}{2}\int ds\:ds'\int dt\:dt'\:i{\hat
R}_{j}(s,t)\int\frac{d^{3}k}{(2\pi)^{3}}k_{j}k_{l}\left|V(k)\right|^{2}S({\bf
k},t-t')\nonumber\\&\:&\qquad\times\exp\left\{i{\bf k}\left[{\bf
R}(s,t)-{\bf R}(s',t')\right]\right\}i{\hat R}_{l}(s',t')\nonumber\\&+&\int
ds\:ds'\int dt\:dt'\:i{\hat
R}_{j}(s,t)\int\frac{d^{3}k}{(2\pi)^{3}}k_{j}k_{l}\left|V(k)\right|^{2}P_{l}({\bf
k},t-t')\nonumber\\&\:&\qquad\times\exp\left\{i{\bf k}\left[{\bf
R}(s,t)-{\bf R}(s',t')\right]\right\}\Bigg\}\label{GF2}
\end{eqnarray}
where the correlation function
\begin{equation}
S({\bf k},t)\equiv\left<\rho({\bf k},t)\rho(-{\bf k},0)\right>_{1}
\end{equation}
and the response function
\begin{equation}
P_{l}({\bf k},t)\equiv\left<\rho({\bf k},t)\Pi_{l}(-{\bf k},0)\right>_{1}
\end{equation}
of the matrix are naturally defined. Going beyond the LRT-approximation would
bring us multi-point correlation and response functions.\\We should stress
that in contrast to the matrix with a quenched disorder which was considered in
\cite{Horner,Cule} in our case the matrix has its own intrinsic dynamical
evolution which is considered as given. For example, for the glass forming
matrix, which is our prime interest here, the correlation and response
functions are assumed to be governed by the G\"otze mode-coupling
equations \cite{Gotze}.
\subsection{The Hartree approximation}
The Hartree approximation (which is actually equivalent to the Feynman
variational principle) was recently used for the replica field theory of
random manifolds \cite{Mezard} as well as for the dynamics of manifolds
\cite{Horner} and sine-Gordon model \cite{Cule} in a random media.

In the Hartree approximation the real MSR-action is replaced by a Gaussian
action in such a way that all terms which include more than two fields
$R_{j}(s,t)$ or/and ${\hat R}_{j}(s,t)$ are written in all possible ways as
products of pairs of $R_{j}(s,t)$ or/and ${\hat R}_{j}(s,t)$, coupled to
selfconsistent averages of the remaining fields.
 As a result the
Hartree-action is a Gaussian functional with coefficients, which could be
represented in terms of correlation and response functions. After these
straightforward calculations (details can be found in the Appendix B) the GF
(\ref{GF2}) takes the form
\begin{eqnarray}
Z\left\{\cdots\right\}&=&\int DR_j(s,t)D{\hat
  R}_j(s,t)\exp\Bigg\{-A_{0}\left[{\bf R}(s,t),{\bf {\hat
R}}(s,t)\right]\nonumber\\&+&\int_{0}^{N}ds\:ds'\int_{-\infty}^{\infty}
dt\int_{-\infty}^{t}dt'\:i{\hat
R}_{j}(s,t)R_{j}(s',t')\lambda(s,s';t,t')\nonumber\\&-&\int_{0}^{N}ds\:ds'\int_{-\infty}^{\infty}
dt\:i{\hat
R}_{j}(s,t)R_{j}(s,t)\int_{-\infty}^{t}dt'\lambda(s,s';t,t')\nonumber\\&+&\frac{1}{2}\int_{0}^{N}ds\:ds'\int_{-\infty}^{\infty}
dt\int_{-\infty}^{t}dt'\:i{\hat
  R}_{j}(s,t)i{\hat R}_{j}(s',t')\chi(s,s';t,t')\Bigg\}\label{Hartree}
\end{eqnarray}
where
\begin{eqnarray}
\lambda(s,s';t,t')&=&\frac{1}{3}G(s,s';t,t')\int\frac{d^{3}k}{(2\pi)^{3}}k^{4}\left|V(k)\right|^{2}F({\bf
  k};s,s';t,t')S({\bf
k};t,t')\nonumber\\&-&\int\frac{d^{3}k}{(2\pi)^{3}}k^{2}\left|V(k)\right|^{2}F({\bf
  k};s,s';t,t')P({\bf k};t,t')\label{lambda}
\end{eqnarray}
and
\begin{eqnarray}
\chi(s,s';t,t')&=&\int\frac{d^{3}k}{(2\pi)^{3}}k^{2}\left|V(k)\right|^{2}F({\bf
  k};s,s';t,t')S({\bf k};t,t')\label{chi}
\end{eqnarray}
In eq.~(\ref{lambda},\ref{chi}) the response function
\begin{equation}
G(s,s';t,t')=\left<i{\bf {\hat R}}(s',t'){\bf R}(s,t)\right>\:\:,
\end{equation}
the density correlator
\begin{equation}
F({\bf
k};s,s';t,t')=\exp\left\{-\frac{k^{2}}{3}Q(s,s';t,t')\right\}\label{dc}
\end{equation}
with
\begin{eqnarray}
Q(s,s';t,t')&\equiv&\left<{\bf R}(s,t){\bf R}(s,t)\right>-\left<{\bf
    R}(s,t){\bf R}(s',t')\right>\nonumber\\
&=&C(s,s;t,t)-C(s,s';t,t')
\end{eqnarray}
and the longitudinal part of the matrix response function
\begin{equation}
P({\bf k};t,t')=ik_{j}P_{j}({\bf k};t,t')
\end{equation}
are defined. The pointed brackets denote the selfconsistent averaging with the
Hartree-type GF (\ref{Hartree}).\\
Up to now we considered the general off-equilibrium dynamics with the only
restriction of causality \cite{Dominicis,Bausch,Ros}. We now assume that for
very large time moments $t$ and $t'$, where the difference $t-t'$ is finite so
that $\frac{t-t'}{t}\rightarrow 0$, time homogenity and
the fluctuation-dissipation theorem (FDT) holds. This implies
\begin{eqnarray}
G(s,s';t,t')=G(s,s';t-t')=\beta\frac{\partial}{\partial
  t'}Q(s,s';t-t')\qquad,\qquad &t>t'&\label{FDT1}\\
P({\bf k};t,t')=P({\bf k};t-t')=\beta\frac{\partial}{\partial
  t'}S({\bf k};t-t')\qquad,\qquad &t>t'&\label{FDT}
\end{eqnarray}
where $\beta\equiv 1/T$.
By using this in eq.~(\ref{Hartree}) and after integration by parts in the
integrals over $t'$ the GF in
Hartree approximation takes the form
\begin{eqnarray}
Z\left\{\cdots\right\}&=&\int DR_j(s,t)D{\hat
  R}_j(s,t)\nonumber\\
&\times&\exp\Bigg\{\int_{0}^{N}ds\:ds'\int_{-\infty}^{\infty} dtdt'\:i{\hat
    R}_{j}(s,t)\Bigg[\xi_{0}\delta(t-t')\delta(s-s')+\nonumber\\
&\:&+\:\Theta(t-t')\beta\int\frac{d^{3}k}{(2\pi)^{3}}k^{2}|V(k)|^{2}F({\bf
    k};s,s';t-t')S({\bf k};t-t')\Bigg]\frac{\partial}{\partial
  t'}R_{j}(s',t')\nonumber\\
&-&\int_{0}^{N}ds\:ds'\int_{-\infty}^{\infty}dt\:i{\hat
  R}_{j}(s,t)\Bigg[\varepsilon\delta(s-s')\frac{\partial^{2}}{\partial
s^{2}}+\beta\int\frac{d^{3}k}{(2\pi)^{3}}k^{2}|V(k)|^{2}S_{st}({\bf
k})\nonumber\\
&\:&\:\times\left[F_{st}({\bf
    k};s,s')-\delta(s-s')\int_{0}^{N}ds^{''}F_{st}({\bf
k};s,s^{''})\right]\Bigg]R_{j}(s',t)\nonumber\\&+&T\int_{0}^{N}ds\:ds'\int_{-\infty}^{\infty}
dtdt'\Bigg[\xi_{0}\delta(t-t')\delta(s-s')+\Theta(t-t')\beta\nonumber\\&\:&\:\int\frac{d^{3}k}{(2\pi)^{3}}k^{2}|V(k)|^{2}F({\bf
    k};s,s';t-t')S({\bf k};t-t')\Bigg]i{\hat R}_{j}(s,t)i{\hat
R}_{j}(s',t')\label{GFh}
\end{eqnarray}
where the subscript $''st''$ indicates the static correlation functions. This generating functional immediately leads to the following generalized
Rouse equation
(GRE)
\begin{eqnarray}
\xi_{0}\frac{\partial}{\partial
t}R_{j}(s,t)&+&\int_{0}^{N}ds'\int_{-\infty}^{t}dt'\Gamma(s,s';t-t')\frac{\partial}{\partial
t'}R_{j}(s',t')\nonumber\\&-&\int_{0}^{N}ds'\Omega(s,s')R_{j}(s',t)={\cal
F}_{j}(s,t)\label{GRE},
\end{eqnarray}
where the memory function
\begin{equation}
\Gamma(s,s';t-t')=\beta\int\frac{d^{3}k}{(2\pi)^{3}}k^{2}|V(k)|^{2}F({\bf
    k};s,s';t-t')S({\bf k};t-t')\label{Gamma}
\end{equation}
and the effective elastic susceptibility
\begin{eqnarray}
\Omega(s,s')&=&\varepsilon\delta(s-s')\frac{\partial^{2}}{\partial
s^{2}}+\beta\int\frac{d^{3}k}{(2\pi)^{3}}k^{2}|V(k)|^{2}S_{st}({\bf
k})\times\nonumber\\&\:&\times\left[F_{st}({\bf
    k};s,s')-\delta(s-s')\int_{0}^{N}ds^{''}F_{st}({\bf
k};s,s^{''})\right]\label{Omega}
\end{eqnarray}
are defined. The correlation function of the random force ${\cal F}_{j}$ is
given by
\begin{equation}
\left<{\cal F}_{i}(s,t){\cal
F}_{j}(s',t')\right>=2T\delta_{ij}\Big[\xi_{0}\delta(s-s')\delta(t-t')+\Theta(t-t')\Gamma(s,s';t-t')\Big]\label{force}
\end{equation}
As a result we have obtained basically the same GRE as in the papers
\cite{Hess,Schweizer1,Schweizer2} but with one additional elastic term. This
term (see the 2-nd term in eq.~(\ref{Omega})) is mainly inversely proportional
to the temperature and is, in contrast to the first term, of an energetic
nature.
The two factors of $kV(k)$ quantify the forces exerted by a pair of
surrounding segments on the test chain segments $s$ and $s'$, whereas the
$S_{st}({\bf k})$ and $F_{st}({\bf k};s,s')$ factors quantify the static
correlations between the segments of surrounding and test chain segments,
respectively. In \cite{Hess,Schweizer1,Schweizer2} only the entropic elastic
part was taken into account.
The memory function (\ref{Gamma}) has the same form as in
\cite{Hess,Schweizer1,Schweizer2} and the relationship (\ref{force}) is
assured as soon as the FDT (\ref{FDT1}) and (\ref{FDT}) is fullfilled.
\subsection{Generalized Rouse equations for the off-equilibrium dynamics}
In this subsection we give GRE's for the more general case when the time
homogenity (stationarity) and the FDT do not hold \cite{14}.\\
By employing the standard way \cite{Horner} one can derive two coupled
equations of motion for correlators $C(s,s';t,t')$ and response
functions $G(s,s';t,t')$
\begin{eqnarray}
\Big[\xi_{0}\frac{\partial}{\partial
    t}&-&\varepsilon\frac{\partial^{2}}{\partial
s^{2}}-\int_{0}^{N}ds^{''}\int_{-\infty}^{t}dt^{''}\lambda(s,s^{''};t,t^{''})\Big]G(s,s';t,t')\nonumber\\&+&\int_{0}^{N}ds^{''}\int_{-t'}^{t}dt^{''}\lambda(s,s^{''};t,t^{''})G(s^{''},s';t^{''},t')=\delta(s-s')\delta(t-t')\label{cor}
\end{eqnarray}
\begin{eqnarray}
\Big[\xi_{0}\frac{\partial}{\partial
    t}&-&\varepsilon\frac{\partial^{2}}{\partial
s^{2}}-\int_{0}^{N}ds^{''}\int_{-\infty}^{t}dt^{''}\lambda(s,s^{''};t,t^{''})\Big]C(s,s';t,t')\nonumber\\&+&\int_{0}^{N}ds^{''}\int_{-\infty}^{t}dt^{''}\lambda(s,s^{''};t,t^{''})C(s^{''},s';t^{''},t')\nonumber\\
&+&\int_{0}^{N}ds^{''}\int_{-\infty}^{t}dt^{''}\chi(s,s^{''};t,t^{''})G(s',s^{''};t',t^{''})=2T\xi_{0}G(s',s;t',t)\label{response}
\end{eqnarray}
with the initial conditions
\begin{eqnarray}
\xi_{0}G(s,s';t=t'+0^{+})&=&\delta(s-s')\nonumber\\
G(s,s';t=t')&=&0\:,\qquad\qquad t\leq t'\label{ini1}
\end{eqnarray}
and
\begin{eqnarray}
C(s,s';t=t')=\left<{\bf R}(s,t){\bf R}(s',t)\right>\label{ini2}
\end{eqnarray}
In the stationary case all correlators and response functions in
eq.~(\ref{cor}-\ref{ini2}) only depend from the differences of time moments,
$t-t'$. If we assume again that FDT (\ref{FDT1}) and (\ref{FDT}) holds, then
from
eq.~(\ref{response}) after performing the integrations by parts (in the
integrals over $t^{''}$) one arrive at the GRE for $t>0$
\begin{eqnarray}
\xi_{0}\frac{\partial}{\partial
t}C(s,s';t)&+&\int_{-\infty}^{t}dt'\int_{0}^{N}ds^{''}\Gamma(s,s^{''};t-t')\frac{\partial}{\partial
t'}C(s^{''},s';t')\nonumber\\
&-&\int_{0}^{N}ds^{''}\Omega(s,s^{''})C(s^{''},s';t)=0\label{Gcor}.
\end{eqnarray}
Of course, eq.~(\ref{Gcor}) could be obtained immediately from eq.~(\ref{GRE})
by multiplying both sides with $R_{j}(s',0)$, averaging and taking into account
that because of causality $\left<{\cal F}(s,t){\bf R}(s',0)\right>=0$ at
$t>0$.
We will use the GRE eq.(\ref{Gcor}), where the
functions $\Gamma$ and $\Omega$
are given by eqs. (\ref{Gamma}, \ref{Omega}), in
the next section for the
investigation of the test chain ergodicity breaking (freezing).
\section{Some statical and dynamical properties of the test chain}
The new features of the GRE (\ref{Gcor}) relative to the standard Rouse
equation are that it contains the integral convolution with respect to the
$s-$variable in the frictional term as well as in the elastic term. The
frictional term is also non-local in time. All these things together should
change the statical and dynamical behaviour of the Gaussian test chain in
comparison to the ideal chain.\\
We also should stress that the GRE is substantially nonlinear because the
memory function (\ref{Gamma}) depends from the test chain correlator
$C(s,s';t)$ in such a way that a positive feedback obviously exists. That
is the reason why one could expect that eq.~(\ref{Gcor})
shows an ergodicity breaking in the spirit of G\"otze's glass transition theory
\cite{Gotze}. \\
As usual it is convenient to introduce the standard Rouse mode variables
\cite{Doi}:
\begin{equation}
{\bf X}(p,t)=\frac{1}{N}\int_{0}^{N}ds{\bf R}(s,t)\cos\left(\frac{p\pi
    s}{N}\right)\label{RT}
\end{equation}
with the inverse transformation
\begin{equation}
{\bf R}(s,t)={\bf X}(p=0,t)+2\sum_{p=1}^{\infty}{\bf
  X}(p,t)\cos\left(\frac{p\pi s}{N}\right).
\end{equation}
In general one also needs a 2-dimensional Rouse transformation
\begin{equation}
\Gamma(p,q;t)=\frac{1}{N^{2}}\int_{0}^{N}ds^{'}\int_{0}^{N}ds^{''}\Gamma(s^{'},s^{''};t)\cos\left(\frac{p\pi
s^{'}}{N}\right)\cos\left(\frac{q\pi s^{''}}{N}\right)\label{2RT}
\end{equation}
where functions like $\Gamma(s',s^{''})$ should be treated like $N\times
N$-matrices. For example the density correlator (\ref{dc}) should be
considered as an exponential function from a $N\times N$-matrix
$Q(s^{'},s^{''})$ and the series expansion holds :
\begin{eqnarray}
F(s,s')&=&1-\frac{k^{2}}{3}Q(s,s^{'})+\frac{1}{2}\left(\frac{k^{2}}{3}\right)^{2}\int_{0}^{N}ds^{''}\:Q(s,s^{''})Q(s^{''},s^{'})\nonumber\\
&-&\frac{1}{3!}\left(\frac{k^{2}}{3}\right)^{3}\int_{0}^{N}ds^{''}\int_{0}^{N}ds^{'''}\:Q(s,s^{''})Q(s^{''},s^{'''})Q(s^{'''},s^{'})+\ldots\label{expa}
\end{eqnarray}
We also assume that matrices in the Rouse mode representation are nearby
diagonal
\begin{eqnarray}
\Gamma(p,q)&=&\delta_{p,q}\Gamma(p)\\
Q(p,q)&=&\delta_{p,q}Q(p)\\
\Omega(p,q)&=&\delta_{p,q}\Omega(p)\label{diag}
\end{eqnarray}
for any $p$ and $q$ not equal zero \cite{Doi}.\\
Then as a result of Rouse mode transformation
the GRE for the Rouse mode time correlation function, $C(p,t)\equiv\left<{\bf
    X}(p,t){\bf X}(p,0)\right>$, takes the form (for $p\neq 0$)
\begin{eqnarray}
\xi_{0}\frac{d}{dt}C(p,t)+\int_{0}^{t}dt'\:\Gamma(p,t-t')\frac{\partial}{\partial
  t'}C(p,t')+\Omega(p)C(p,t)=0\label{GREr}
\end{eqnarray}
where
\begin{equation}
\Gamma(p,t)=\beta\int\frac{d^{3}k}{(2\pi)^{3}}k^{2}|V(k)|^{2}\left\{\exp\left[\frac{k^{2}}{3}NC(p,t)\right]-1\right\}S({\bf
  k},t)\label{Gammap}
\end{equation}
and
\begin{equation}
\Omega(p)=\varepsilon\left(\frac{p\pi}{N}\right)^{2}-\beta
N\int\frac{d^{3}k}{(2\pi)^{3}}k^{2}|V(k)|^{2}S_{st}({\bf k})\Big[F_{st}({\bf
  k};p)-F_{st}({\bf k};p=0)\Big]\label{Omegap}
\end{equation}
For $p=0$ the GRE describes the dynamics of the centre of mass
\begin{equation}
{\bf R}_{c.m}(t)\equiv{\bf X}(p=0,t)=\frac{1}{N}\int_{0}^{N}ds\:{\bf R}(s,t)
\end{equation}
and has the following form
\begin{eqnarray}
\xi_{0}\frac{d}{dt}{\bf
R}_{c.m}(t)&+&\beta\int_{0}^{t}dt'\int\frac{d^{3}k}{(2\pi)^{3}}k^{2}|V(k)|^{2}F({\bf
  k};p=0,q=0,t-t')\nonumber\\&\:&\qquad\times S({\bf k},t-t')
\frac{d}{dt'}{\bf
  R}_{c.m}(t')={\bf f}_{c.m}(t)\label{c.m}
\end{eqnarray}
with
\begin{equation}
\left(f_{c.m}\right)_{j}(t)\equiv\frac{1}{N}\int_{0}^{N}ds\:{\cal F}_{j}(s,t)
\end{equation}
and
\begin{equation}
F({\bf
k};p=0,q=0;t)=\frac{1}{N^{2}}\int_{0}^{N}ds'\int_{0}^{N}ds^{''}\exp\left\{-\frac{k^{2}}{3}Q(s',s^{''};t)\right\}\label{47}
\end{equation}
As a result all Rouse mode variables relax independently. The conclusion that
the Rouse modes are still "good eigenmodes" even in the melt is supported by
Monte-Carlo \cite{Binder} and molecular-dynamic \cite{15} simulations.\\
For cases where the assumption of diagonality [38,39,\ref{diag}] cannot be justified,
the Rouse modes do not decouple and one have to go back to eq.~(\ref{Gcor}). In
the Rouse mode representation it reads as
\begin{eqnarray}
\xi_{0}\frac{d}{dt}C(p,t)+\int_{0}^{t}dt'\int dq\:\Gamma(p,q;t-t')\frac{\partial}{\partial
  t'}C(q,t')+\int dq\:\Omega(p,q)C(q,t)=0\label{GRErq}
\end{eqnarray}
\subsection{Static properties}
As we have already discussed in sec.II.B the interaction with the
surrounding segments renormalizes the elastic properties of the Rouse chain so
that the test chains elastic susceptibility is given by eq.~(\ref{Omegap}). The
additional elastic term in GRE leads to the renormalized static normal modes
correlator
\begin{equation}
C_{st}(p)=\frac{T}{2N\Omega(p)}\label{Cst}
\end{equation}
Explicit evaluation of the $\Omega(p)$ can be done if we use for the static
correlator $F_{st}({\bf k};p)$ the standard Rouse expression
\begin{equation}
F_{st}({\bf
k};p)=\frac{1}{N}\int_{0}^{N}ds\exp\left(-\frac{l^{2}k^{2}s}{6}\right)\cos\left(\frac{p\pi
s}{N}\right).
\end{equation}
Then the calculation yields for the two limiting cases
\begin{eqnarray}
\Omega(p)=\left\{
\begin{array}{r@{\quad:\quad}l}\left(\frac{p\pi}{N}\right)^{2}\left[\varepsilon+\frac{\beta}{2\pi^{2}}\int_{0}^{l^{-1}}dk\:k^{4}|V(k)|^{2}S_{st}(k)g(k^{2}R_{g}^{2})\right]+{\cal
O}(p^{4}) & \frac{p\pi}{N}\ll 1\quad (50a) \\
\left(\frac{p\pi}{N}\right)^{2}\varepsilon+\frac{\beta}{4\pi^{2}}\int_{0}^{l^{-1}}dk\:k^{4}|V(k)|^{2}S_{st}(k)\left(\frac{6}{l^{2}k^{2}}\right)\left(1-e^{-k^{2}R_{g}^{2}}\right)
&  \frac{p\pi}{N}\simeq 1\quad\: (50b)\end{array}\right.\nonumber
\end{eqnarray}
\stepcounter{equation}
where
\begin {equation}
g(x)=\frac{1}{x^{3}}\left[2-(x^{2}+2x+2)e^{-x}\right],\qquad
R_{g}=\frac{Nl^{2}}{6}
\end{equation}
and we have chosen $l^{-1}$ as a cutting parameter. It is evident from the
previous eqs. (50$a$, 50$b$) that at small $p$
\begin{itemize}
\item the elastic modulus gains an energetic component which, in contrast to
the
entropic part $\varepsilon$, increase with the cooling of the system,
\item initially absolutely flexible chains acquires a stiffness because of
  terms of order $p^{4}$ and higher.
\end{itemize}
At large $p$ the elastic behaviour
  reduces to the standard Rouse one, as it is expected. In Fig.1 is shown the
  result of a numerical calculation of the static correlator (\ref{Cst}). The
  Fourier component of the potential is taken, as it is customary e.g. in the
  theory of neutron scattering \cite{16}, in the form of a pseudo potential
  approximation, $V(k)=\gamma\sigma^{3}$, where $\gamma$ and $\sigma$ have
  dimensions of molecular energy and distance, respectively. The static
  structure factor $S_{st}({\bf k})$ is chosen in the form of the
  Percus-Yevick's simple liquid model \cite{17}. One can see that
  for $N=500$ the small Rouse mode index limit (50$a$) starts at
$\frac{p}{N}\leq
  3\cdot 10^{-3}$
  whereas the opposit limit (50$b$) is fullfilled at $\frac{p}{N}\geq
10^{-1}$. Because the
  correlator $C_{st}(p)$ depends mainly from $p/N$, for relatively short
  test chains the high mode index limit (50$b$) is shifted into the window of
  calculations (see Fig.1 for N=20).\\
At least qualitatively this deviation from the standard Rouse behaviour have
been seen by Kremer and Grest in their MD-simulations (see Fig.3 in \cite{15}).
\subsection{The test chain ergodicity breaking in a glass forming \\matrix}
First we consider the case $p\neq 0$. In the nonergodic state the Rouse mode
correlation functions can be represented as
\begin{equation}
\Psi(p,t)\equiv\frac{C(p,t)}{C_{st}(p)}=\Psi_{reg}(p,t)+g(p)
\end{equation}
where the non-ergodicity parameter
\begin{equation}
g(p)\equiv \lim_{t\rightarrow \infty}\Psi(p,t)\label{Psi}
\end{equation}
was introduced and $\Psi_{reg}(p,t\rightarrow \infty)=0$.

For the correlation function of the glassy matrix we can use the standard
result of the glass transition theory \cite{Gotze}
\begin{equation}
\phi({\bf k},t)\equiv\frac{S({\bf k},t)}{S_{st}({\bf k})}=f^{c}({\bf
  k})+h({\bf
  k})\Delta^{1/2}\left(\frac{\tau_{\Delta}}{t}\right)^{a}\label{Gg}
\end{equation}
where the proximity parameter $\Delta\equiv (T_{G}-T)/T_{G}$ is defined and
$T_{G}$ is the temperature of the matrix ergodicity breaking (G\"otze
temperature). In eq.~(\ref{Gg}) $f^{c}({\bf
  k})$ is the non-ergodicity parameter of the matrix,
$\tau_{\Delta}\propto \Delta^{-1/2a}$ is the characteristic time scale, $a$
is the characteristic exponent, $0<a<1/2$ and $h({\bf k})$ is some amplitude.\\
In order to derive the equation for $g(p)$ let us take the limit $t\rightarrow
\infty$ in eq.~(\ref{GREr}) keeping in mind the definitions (\ref{Psi})
and (\ref{Gg}). Very close to the test chain ergodicity breaking temperature
$T_{c}(p)$, $g(p)$ goes to zero (A-type transition \cite{Gotze}) and we can
expand the exponential function in eq.~(\ref{Gammap}) up to the first order
with respect to $g(p)$. The solution of the resulting equation has the simple
form
\begin{equation}
g(p)=1-\frac{6\Omega(p)^{2}}{\int\frac{d^{3}k}{(2\pi)^{3}}k^{4}|V(k)|^{2}S_{st}({\bf
  k})f^{c}({\bf k})}\label{Tp}
\end{equation}
The critical temperature $T_{c}(p)$ is determined by the equation
\begin{equation}
g(p,T=T_{c})=0\label{Tc}
\end{equation}
The numerical solution of eq.~(\ref{Tc}) is given in Fig.2. It is obviously
that if the entropic part of $\Omega(p)$ dominates, the critical temperature
is given by
\begin{equation}
T_{c}(p)\propto \left(\frac{N}{\pi p}\right)^{2}\label{pq}.
\end{equation}
Fig.2 really shows that this law (\ref{pq}) is well satisfied due to the
fact that the critical temperatures $T_{c}(p)$ are quite high. But for low
temperatures the energetic contribution in $\Omega(p)$ is enhanced which leads
to a deviation from this simple $(N/p)^{2}$-dependence.\\
Now we consider the case for $p=0$. The equation (\ref{c.m}) for the velocity
of the center of mass
\begin{equation}
{\bf v}(t)\equiv\frac{d}{dt}{\bf R}_{c.m}(t)
\end{equation}
leads to the equation for the velocity correlator
\begin{equation}
\xi_{0}\left<v_{j}(t)v_{i}(0)\right>+\int_{0}^{t}dt'\Gamma(t-t')\left<v_{j}(t')v_{i}(0)\right>=\left<(f_{c.m})_{j}(t)v_{i}(0)\right>\label{c.m.c}
\end{equation}
where
\begin{equation}
\Gamma(t)=\beta\int\frac{d^{3}k}{(2\pi)^{3}}k^{2}|V(k)|^{2}F({\bf
  k};p=q=0;t)S({\bf k},t)\label{Gammat}
\end{equation}
Because of causality the correlator on the r.h.s. of eq.~(\ref{c.m.c}) has the
form
\begin{eqnarray}
\left<(f_{c.m})_{j}(t)v_{i}(0)\right>=\left\{\begin{array}{r@{\quad,\quad}l}0
& t>0\\
\neq 0 & t=0\end{array}\right.
\end{eqnarray}
where, as it comes from eq.~(\ref{c.m})
\begin{equation}
v_{i}(0)=\frac{1}{\xi_{0}}\left(f_{c.m}\right)_{i}(0)\label{f0}
\end{equation}
Taking into account the definition of $\left(f_{c.m}\right)_{i}(t)$ and
eq.~(\ref{force}) this yields to the correlator
\begin{eqnarray}
\left<(f_{c.m})_{j}(t)v_{i}(0)\right>&=&
2T\delta_{ij}\delta(t)\frac{1}{N^{2}}\int_{0}^{N}ds'\int_{0}^{N}ds^{''}\delta(s'-s^{''})\nonumber\\
&=&\frac{2T}{N}\delta_{ij}\delta(t)\label{lk}
\end{eqnarray}
Because of the causality property (61) only the $\delta$-functional
term on the r.h.s. of eq.~(\ref{force}) contributes to the correlator
(\ref{lk}). Therefore the resulting equation for the self-diffusion
coefficient
\begin{equation}
D\equiv\frac{1}{3}\int_{0}^{\infty}dt\left<{\bf v}(t){\bf v}(0)\right>
\end{equation}
takes the form
\begin{equation}
D=\frac{T}{N\left[\xi_{0}+\int_{0}^{\infty}dt\Gamma(t)\right]}\label{D}
\end{equation}
which was obtained before in \cite{Hess,Schweizer1}.\\
One can calculate the second term in the denominator of eq.~(\ref{D})
selfconsistently. Because now the relevant times $t\gg \tau_{rouse}$ the
approximation
\begin{equation}
Q(s',s^{''};t)=6Dt+l^{2}|s'-s^{''}|+{\mbox{const.}}\label{s.r}
\end{equation}
could be used in eq.~(\ref{c.m}). Then the density correlator (\ref{47}) is
given by
\begin{equation}
F({\bf
k};p=q=0;t)=\frac{1}{\frac{k^{2}l^{2}}{12}+N^{-1}}\exp\left(-k^{2}Dt\right)\label{De}
\end{equation}
With the use of eqs.~(\ref{De}),(\ref{Gg}) and eq.~(\ref{Gammat}) in the limit
$D\rightarrow 0$ eq.~(\ref{D}) becomes
\begin{equation}
D=\frac{T}{N\left[\xi_{0}+\frac{1}{TD}\int\frac{d^{3}k}{(2\pi)^{3}}\frac{|V(k)|^{2}S_{st}({\bf
        k})}{\frac{k^{2}l^{2}}{12}+N^{-1}}f({\bf k)}\right]}\label{Del}
\end{equation}
where the denominator is given by static properties only.
Similar statements have been suggested already in \cite{vil1,vil2}
The solution of eq.~(\ref{Del}) has the simple form
\begin{equation}
D=D_{0}\left(1-\frac{N}{T^{2}}\chi\right)\label{LD}
\end{equation}
where
\begin{equation}
\chi=\int\frac{d^{3}k}{(2\pi)^{3}}\frac{|V(k)|^{2}S_{st}({\bf
        k})}{\frac{k^{2}l^{2}}{12}+N^{-1}}f({\bf k})\label{chi2}
\end{equation}
Finally, the temperature of the ergodicity breaking (localization) for the
mode $p=0$ of the test chain is
\begin{equation}
T_{c}(p=0)=\left(N\chi\right)^{1/2}
\end{equation}
Fig.3 shows the results of numerical calculations of $T_{c}(p=0)$ and
$T_{c}(p=1)$ as functions of $N$. One can see that in the reasonable range of
parameters $T_{c}(p=0)>T_{c}(p=1)$. As a result one can say that on cooling of
a test chain in a glassy matrix the mode $p=0$ is the first to be freezed. On
the subsequent cooling the modes $p=1,2,\ldots,N$ are freezed successively,
\begin{equation}
T_{G}>T_{c}(p=0)>T_{c}(p=1)>T_{c}(p=2)>\ldots T_{c}(p=N).
\end{equation}
It is apparent that the system studied here is a nontrivial polymeric
generalization of the model introduced by Sj\"ogren \cite{18}. This model was
used for the investigation of the $\beta$-peak in the spectrum of glass forming
systems \cite{19}.
\section{Discussion}
In this paper we have derived a GRE for a test
polymer chain in a polymer (or
non-polymer) matrix which has its own intrinsic dynamics, e.g. the glassy
dynamics \cite{Gotze}.
We have used here the MSR-functional integral technique which could be
considered as an alternative to the projection operator formalism
\cite{Schweizer1}. One of the difficulties in this formalism
is the necessity
of dealing with the projected dynamic, which is difficult to
handle with
explicitly. On the contrary in MSR-technique the dynamic of slow variables is
well defined and several approximations which one have to employ could be
justified.

In the interaction of the test chain with the surrounding matrix only two-point correlation and response functions are involved. In terms of
MCA \cite{Schweizer1} this obviously corresponds to the
projection of the
generalized forces only onto the bilinear variables:
product of test chain density and matrix density.

To handle with the action in the GF of
the test chain we used the
Hartree-type approximation (i.e., equivalent to the
Feynman variational
principle) \cite{Horner,Cule,Mezard}, which is reasonable when the
fluctuations of the test chain are Gaussian.
In the case of a polymer melt (high densitiy) this
is indeed the case due to the screening effects for the
excluded volume \cite{Doi}.

The use of the Hartree-type makes the problem that we deal with analytically
amenable and results in the GRE's for the case when the FDT holds as well as
for
the case when FDT does not hold. In this paper we have restricted ourselves to
the first case and have shown that the interaction with the matrix
renormalizes not only the friction coefficient (which makes the chain
non-Markovian) but also the elastic modulus (which changes the static
correlator). The form of the static correlator for the Rouse mode variables is
qualitatively supported by MD-simulations \cite{15}.

As regards the dynamical behaviour, we have shown that the test chain in a
glassy matrix (with the matrix glass transition temperature $T_{G}$) undergoes
the ergodicity breaking transition at a temperature $T_{c}(p)\leq T_{G}$. The
critical temperature $T_{c}(p)$ could be parametrized with the Rouse mode
index $p$ and is a decreasing function of $p$.

We have considered only the A-type transition which is assured by the bilinear term in the expansion of eq.~(\ref{Gammap}). It seems reasonable that keeping the whole exponential function in eq.~(\ref{Gammap}) might lead to a B-type transition also. The results also essentially would change if the off-diagonal elements in the matrix (\ref{2RT}) can not be neglected (see eq.~(\ref{GRErq})). In this case only one ideal transition temperature $T_c$ would be possible. The general theory of a A-type transition was discussed in \cite{24}.

This picture of freezing here should not be mixed with a different one, the
underlying glass transition by itself (e.g. the glass transition of the matrix
at $T=T_{G}$). According to the present view of this phenomenon \cite{Gotze},
the spontaneous arrest of the density fluctuations is driven by those of the
microscopic lengthscale $k_{0}$, where $k_{0}$ is the wave vector which
corresponds to the structure factor's main maximum. The freezing of these
fluctuations then arrests the others through the mode coupling.

\acknowledgments
Two of us gratefully acknowledge support from the Deutsche
Forschungsgemeinschaft
through the Sonderforschungsbereich 262 (V.G.Rostiashvili) and the
Bundesministerium f\"ur Bildung und Forschung (M.Rehkopf) for financial
support.
We also greatly acknowledge helpful discussions with
J.Baschnagel, K.Binder, K.Kremer, W.G\"otze and R.Schilling.
\begin{appendix}
\section{Response field density Correlator}
It is more convenient to handle with the spacial Fourier transformation of
this correlator
\begin{eqnarray}
&{\:}&\left<\Pi_{l}({\bf k},t)\Pi_{j}({\bf
-k},0)\right>_{1}\label{Ap}\\&=&\sum_{(p,m=1)}^{M}\int
dsds'\left<i{\hat r}_{l}^{(p)}(s,t)i{\hat
    r}_{j}^{(m)}(s',0)\exp\left\{i{\bf k}\left[{\bf r}^{(p)}(s,t)-{\bf
        r}^{(m)}(s',0)\right]\right\}\right>_{1}\nonumber\\
&=&\sum_{a,b=0}^{\infty}\frac{1}{a!b!}\sum_{p,m=1}^{M}\int ds
ds'\left<i{\hat r}_{l}^{(p)}(s,t)i{\hat r}_{j}^{(m)}(s',0)[i{\bf k}{\bf
    r}^{(p)}(s,t)]^{a}[-i{\bf k}{\bf r}^{(m)}(s',0)]^{b}\right>_{1}\nonumber
\end{eqnarray}
Such multi-point cumulant response functions (MRF) were considered in
\cite{Ros}. The causality condition for these functions asserts that the time
argument of at least one ${\bf r}$-variable should be the latest one,
otherwise this MRF equals zero. Because of the same reason self-loops of
response functions vanish \cite{Dominicis,Bausch}. MRF's which consists only of
${\hat r}$-variables also vanish.\\
In the case (\ref{Ap}) all time arguments of the ${\bf r}$-variables are equal
to the corresponding time arguments of ${\hat r}$-variables and as a result
the MRF in eq.~(\ref{Ap}) vanishes.
\section{Derivation of the Hartree-type GF}
In order to calculate the bilinear Hartree action, we follow the way mentioned
in sec.II.B. With these strategy in mind the 2-nd term in the
exponent (\ref{GF2}) is evaluated as
\begin{eqnarray}
&\:&\frac{1}{2}\int_{0}^{N}ds ds'\int_{-\infty}^{\infty}dt dt'\:i{\hat
  R}_{j}(s,t)R_{j}(s',t')I_{1}(s,s';t,t')S({\bf k};t,t')\nonumber\\
&+&\frac{1}{2}\int_{0}^{N}ds ds'\int_{-\infty}^{\infty}dt dt'\:i{\hat
  R}_{j}(s',t')R_{j}(s,t)I_{2}(s,s';t,t')S({\bf k};t,t')\nonumber\\
&+&\frac{1}{2}\int_{0}^{N}ds ds'\int_{-\infty}^{\infty}dt dt'\:i{\hat
  R}_{j}(s,t)R_{j}(s,t)I_{3}(s,s';t,t')S({\bf k};t,t')\nonumber\\
&+&\frac{1}{2}\int_{0}^{N}ds ds'\int_{-\infty}^{\infty}dt dt'\:i{\hat
  R}_{j}(s',t')R_{j}(s',t')I_{4}(s,s';t,t')S({\bf k};t,t')\nonumber\\
&+&\frac{1}{2}\int_{0}^{N}ds ds'\int_{-\infty}^{\infty}dt dt'\:i{\hat
  R}_{j}(s,t){\hat R}_{j}(s',t')I_{5}(s,s';t,t')S({\bf
  k};t,t')\nonumber\\
&+&\frac{1}{2}\int_{0}^{N}ds ds'\int_{-\infty}^{\infty}dt
dt'\:iR_{j}(s,t)R_{j}(s',t')I_{6}(s,s';t,t')S({\bf k};t,t')\label{Ap2}
\end{eqnarray}
where
\begin{eqnarray}
I_{1}(s,s';t,t')&\equiv&\left<\frac{\delta}{\delta R_{j}(s',t')}\int
\frac{d^{3}k}{(2\pi)^{3}}k_{j}k_{l}|V(k)|^{2}\exp\left\{i{\bf k}\left[{\bf
      R}(s,t)-{\bf R}(s',t')\right]\right\}i{\hat
  R}_{l}(s',t')\right>\nonumber\\
I_{2}(s,s';t,t')&\equiv&\left<\frac{\delta}{\delta R_{j}(s,t)}\int
\frac{d^{3}k}{(2\pi)^{3}}k_{j}k_{l}|V(k)|^{2}\exp\left\{i{\bf k}\left[{\bf
      R}(s,t)-{\bf R}(s',t')\right]\right\}i{\hat
  R}_{l}(s,t)\right>\nonumber\\
I_{3}(s,s';t,t')&\equiv&\left<\frac{\delta}{\delta R_{j}(s,t)}\int
\frac{d^{3}k}{(2\pi)^{3}}k_{j}k_{l}|V(k)|^{2}\exp\left\{i{\bf k}\left[{\bf
      R}(s,t)-{\bf R}(s',t')\right]\right\}i{\hat
  R}_{l}(s',t')\right>\nonumber\\
I_{4}(s,s';t,t')&\equiv&\left<\frac{\delta}{\delta R_{j}(s',t')}\int
\frac{d^{3}k}{(2\pi)^{3}}k_{j}k_{l}|V(k)|^{2}\exp\left\{i{\bf k}\left[{\bf
      R}(s,t)-{\bf R}(s',t')\right]\right\}i{\hat
  R}_{l}(s,t)\right>\nonumber\\
I_{5}(s,s';t,t')&\equiv&\left<\int\frac{d^{3}k}{(2\pi)^{3}}k_{j}k_{j}|V(k)|^{2}\exp\left\{i{\bf
k}\left[{\bf
      R}(s,t)-{\bf R}(s',t')\right]\right\}\right>\nonumber\\
I_{6}(s,s';t,t')&\equiv&\Bigg<i{\hat
  R}_{j}(s,t)i{\hat
  R}_{l}(s,t)\frac{\delta^{2}}{\delta R_{n}(s,t)\delta
  R_{n}(s',t')}\nonumber\\
&\:&\times\int
\frac{d^{3}k}{(2\pi)^{3}}k_{j}k_{l}|V(k)|^{2}\exp\left\{i{\bf k}\left[{\bf
      R}(s,t)-{\bf R}(s',t')\right]\right\}\Bigg>.\label{Il}
\end{eqnarray}
The pointed brackets in eq.~(\ref{Il}) represent the selfconsistent averaging
with the Gaussian Hartree action. Taking this into account and using the
generalized Wick theorem \cite{13}, after straightforward algebra, we have
\begin{eqnarray}
I_{1}(s,s';t,t')&=&\frac{1}{3}G(s,s';t,t')\int\frac{d^{3}k}{(2\pi)^{3}}k^{4}|V(k)|^{2}\exp\left\{-\frac{k^{2}}{3}Q(s,s';t,t')\right\}\qquad
t>t'\nonumber\\
I_{2}(s,s';t,t')&=&\frac{1}{3}G(s,s';t',t)\int\frac{d^{3}k}{(2\pi)^{3}}k^{4}|V(k)|^{2}\exp\left\{-\frac{k^{2}}{3}Q(s,s';t,t')\right\}\qquad
t'>t\nonumber\\
I_{3}(s,s';t,t')&=&-\frac{1}{3}G(s,s';t,t')\int\frac{d^{3}k}{(2\pi)^{3}}k^{4}|V(k)|^{2}\exp\left\{-\frac{k^{2}}{3}Q(s,s';t,t')\right\}\qquad
t>t'\nonumber\\
&=&-I_{1}(s,s';t,t')\nonumber\\
I_{4}(s,s';t,t')&=&-\frac{1}{3}G(s,s';t',t)\int\frac{d^{3}k}{(2\pi)^{3}}k^{4}|V(k)|^{2}\exp\left\{-\frac{k^{2}}{3}Q(s,s';t,t')\right\}\qquad
t'>t\nonumber\\
&=&-I_{2}(s,s';t,t')\nonumber\\
I_{5}(s,s';t,t')&=&\int\frac{d^{3}k}{(2\pi)^{3}}k^{4}|V(k)|^{2}\exp\left\{-\frac{k^{2}}{3}Q(s,s';t,t')\right\}\nonumber\\
I_{6}(s,s';t,t')&=&0\label{I2}
\end{eqnarray}
where the last equation comes from the fact that the response function
$G(t,t')\propto\Theta(t-t')$ and $G(t',t)\propto\Theta(t'-t)$.\\
The 3-rd term in the exponent eq.~(\ref{GF2}) can be handled in the same
way. The response function for the isotropic matrix has the form
\begin{equation}
P_{j}({\bf k},t)=-\frac{ik_{j}}{k^{2}}P({\bf k},t)
\end{equation}
where $P({\bf k},t)$ is the longitudinal part of the matrix response function.
Then the Hartree approximation of the 3-rd term in the exponent (\ref{GF2})
takes the form
\begin{eqnarray}
\int_{0}^{N}dsds'\int_{-\infty}^{\infty}dt\int_{-\infty}^{t}dt'\left\{i{\hat
    R}_{j}(s,t)R_{j}(s',t')J_{1}(s,s';t,t')+i{\hat
    R}_{j}(s,t)R_{j}(s,t)J_{2}(s,s';t,t')\right\}\nonumber
\end{eqnarray}
where
\begin{eqnarray}
J_{1}(s,s';t,t')&=&-\int\frac{d^{3}k}{(2\pi)^{3}}k^{2}|V(k)|^{2}P({\bf
k};t,t')\exp\left\{-\frac{k^{2}}{3}Q(s,s';t,t')\right\}\nonumber\\
J_{2}(s,s';t,t')&=&\int\frac{d^{3}k}{(2\pi)^{3}}k^{2}|V(k)|^{2}P({\bf
  k};t,t')\exp\left\{-\frac{k^{2}}{3}Q(s,s';t,t')\right\}\nonumber\\
&=&-J_{1}(s,s';t,t')\label{J}
\end{eqnarray}
Taking into account eq.~(\ref{Ap2}) with eq.~(\ref{Il}) and eq.~(\ref{J})
leads to the Hartree-type approximation (\ref{Hartree}).

\end{appendix}

\begin{figure}
\epsfig{file=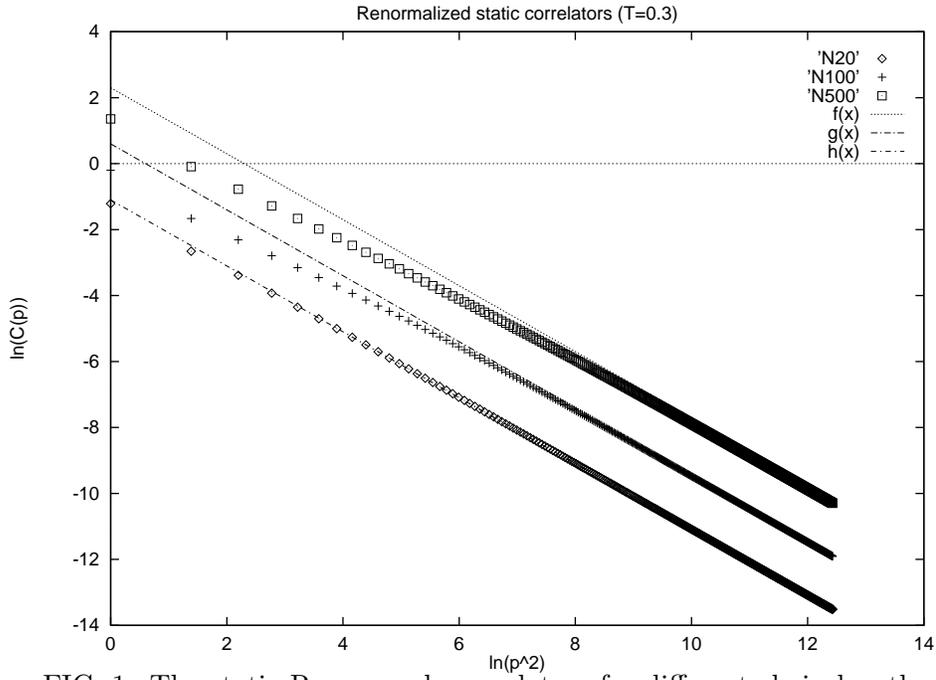}
\caption{The static Rouse mode correlators for different chain lengths. The
lines represents the simple Rouse case. The temperatures are measured in
units of the interaction potential with $\gamma=\sigma=1$.}
\end{figure}
\newpage
\begin{figure}
\epsfig{file=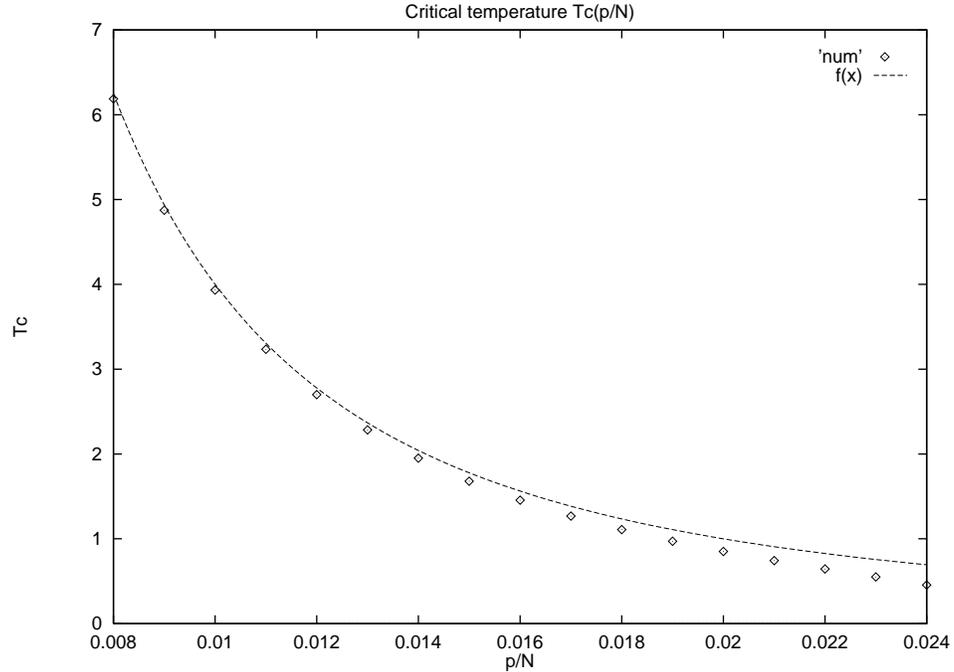}
\caption{The freezing temperatures of the Rouse mode correlators $C(p)$ for
different wavevectors $p/N$, where the temperatures are measured in units of
the interaction potential with $\gamma=\sigma=1$. The dashed line represents
the freezing temperatures, when only the entropic contributions to the
elastic susceptibility are taken into account.}
\end{figure}
\newpage
\begin{figure}
\epsfig{file=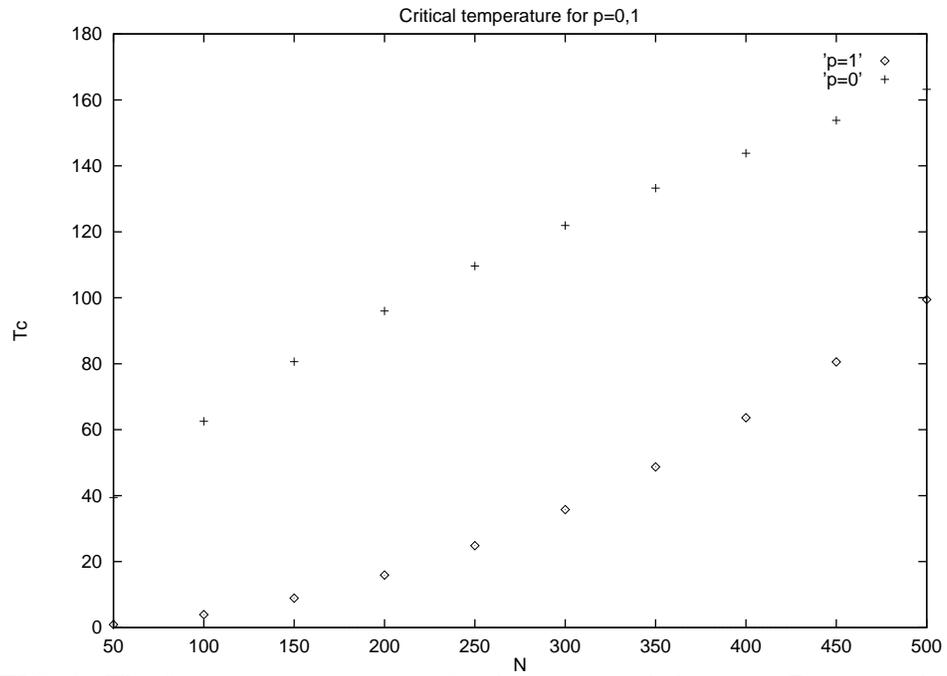}
\caption{The freezing temperatures for the $p=0$ and the $p=1$ Rouse mode
correlators $C(p)$ of the test chain.}
\end{figure}
\end{document}